\documentstyle[prl,aps,epsf,multicol]{revtex}
\begin{document}
\draft
\title{The solid-liquid interfacial free energy of close-packed metals: hard spheres and the Turnbull coefficient}
\author{Brian B. Laird}
\address{Department of Chemistry,
        University of Kansas, Lawrence, Kansas 66045, USA}
%\date{to appear in the {\it Journal of Chemical Physics}, 2001}
\maketitle
\begin{abstract}
% 			    ABSTRACT
Largely due to its role in nucleation and crystal-growth, the free energy of 
the crystal-melt interfacial free energy is an object of considerable interest 
across a number of scientific disciplines, especially in the materials-, 
colloid- and atmospheric sciences. Over fifty years ago, Turnbull observed 
that the interfacial free energies (scaled by the mean interfacial area per 
particle) of a variety of metallic elements exhibit a linear correlation with
the enthalpy of fusion. This correlation provides an important empirical 
"rule-of-thumb" for estimating interfacial free energies, but lacks a 
compelling physical explanation.  In this work we show that the  interfacial 
free energies for close-packed metals are linearly correlated with the melting 
temperature, and are therefore primarily entropic in origin.  We also show that 
the slope of this linear relationship can be determined with quantitative 
accuracy using a hard-sphere model, and that the correlation with the enthalpy 
of fusion reported by Turnbull follows as a consequence of the fact that the 
entropy of fusion for close-packed metals is relatively constant. 
\end{abstract}
\ifpreprintsty
\newpage
\else
\begin{multicols}{2}
\fi

% 			BODY OF THE PAPER
The crystal-melt interfacial free energy, $\gamma$, defined as the reversible work 
required to form a unit area of interface between a crystal and its coexisting 
fluid, plays a central role in determining the kinetics of crystal nucleation 
and growth\cite{Woodruff73,Howe97}.   Unfortunately, direct experimental determinations of this 
quantity are difficult and exist for only a handful of materials. For most 
materials, knowledge of the interfacial free energy  is obtained indirectly 
from measurements of crystal nucleation rates from undercooled fluids, from 
which $\gamma$  is determined within the approximations of classical nucleation 
theory. 

In a seminal 1950 paper\cite{Turnbull50}, Turnbull reported values of the 
crystal-melt interfacial free energy, $\gamma$,  for a variety of materials, mostly 
metallic elements, which were obtained indirectly from nucleation rate 
experiments. For systems in which directly determined values exist with which 
to compare, the values so obtained typically are accurate within about 10-20\%. 
(For example, the interfacial free energy for bismuth was determined\cite{Glicksman69} 
using grain boundary angles to be 61.3 $\times 10^{-3}$ J m$^{-2}$, as compared to the 
value of 54.4 $\times 10^{-3}$ J m$^{-2 }$ obtained from nucleation rate data).  In order 
to compare the results for various systems, Turnbull defined a "gram-atomic" (or molar) 
interfacial free energy  as the free energy of an interface (one atom thick) containing 
Avagadro's number, $N_A$, of atoms (or molecules):
\begin{equation}
\hat{\gamma} = \gamma \rho^{-2/3} N_A \;. 
\end{equation}
The data for $\hat{\gamma}$ was found to exhibit a strong correlation with the latent heat 
of fusion.  Empirically, Turnbull found
\begin{equation}
\hat{\gamma} =  C_T \Delta_{fus} H \; , 
\end{equation}
where the Turnbull coefficient, $C_T$, was found to be approximately 0.45 for metals 
(especially close-packed metals) and 0.32 for many nonmetals.  

There have been attempts to explain this empirical result through the analysis of 
simple models for the structure of the interface\cite{Skapski56,Spaepen75,Spaepen76}, 
but the results are quite sensitive to 
the nature of the assumed model. Inherent in these early models are two basic assumptions: 
{\it 1)} the solid-liquid free energy is primarily entropic in origin, and {\it 2)} the surface free 
energy is due to the increase in entropy associated with the enhanced structure of the 
liquid at the interface (i.e., the crystal is unchanged up to the interface). The first 
assumption is reasonable given the dominant role that packing considerations play in 
determining the structure and thermodynamics of simple liquids\cite{Hansen86} and implies that a 
hard-sphere model should be adequate to describe the interfacial system. However, the 
assumption 2) is at variance with molecular-dynamics simulations of crystal-melt interfaces 
of simple systems\cite{Broughton86,Davidchack98} that show significant structural relaxation 
(evidenced by an increase in the mean-squared displacement from the lattice sites) occurs in 
the crystal as the interface is approached.  Therefore, any theory of the crystal-melt 
interfacial free energy must include a realistic description of both the solid and fluid in 
the interfacial region. 

Recently, we have determined via molecular-dynamics computer simulation, the 
structure\cite{Davidchack98}
and solid-liquid interfacial free energy\cite{Davidchack00} for a system of hard-spheres. 
For this system, which freezes into a face-centered-cubic (fcc) crystal structure, 
$\gamma$ was determined to be slightly anisotropic with an orientationally averaged value of 
\begin{equation}
\gamma_{hs}  = 0.61 \; kT_m /\sigma^2  \; , 
\end{equation}
where $k$ is Boltzmann's constant, $T_m$, is the melting temperature and $\sigma$ is the 
hard-sphere diameter. (The scaling with $k T_m$ is a consequence of the fact that the phase 
behavior in any hard-core systems is purely entropic.)  This value is consistent with a 
value of 0.55$\pm 0.02 \;k T_m/\sigma^2$ obtained from an analysis\cite{Marr94} 
(using nucleation 
theory) of the experimental crystallization kinetics of silica spheres, a system well 
described by a hard-sphere model. Note that, $\gamma_{hs}$ is considerably lower that that 
obtained for a fluid at a structureless hard wall, which was recently calculated\cite{Heni99} 
at the melting density to be 1.99$\pm 0.18 \;k T_m/\sigma^2$, indicating that the entropy 
increase due to the relaxation of the crystal structure near the interface plays an important 
role in determining the  interfacial thermodynamics. 

For the hard-sphere system, the density of the solid at freezing is approximately 
$\rho\sigma^3$ = 1.04, independent of temperature\cite{Hoover68}, therefore Turnbull's gram-atomic interfacial 
free energy can be written
\begin{equation}
\hat{\gamma}_{hs}/R  = 0.59 (0.54)  T_m  \ ,
\end{equation}
where $R$ is the gas constant and the value in parenthesis is that obtained using the value 
of $\gamma_{hs}$ determined 
from nucleation experiments\cite{Marr94}. Eq. 4 suggests that if the interfacial free energy 
of fcc-forming materials is well described by a hard-sphere model, then one should see a 
linear correlation between $\hat{\gamma}/R$ and the melting temperature with a slope in 
the range 0.5 to 0.6.  

\ifpreprintsty
\else
%           FIGURE 1: Correlation of gamma with T_m
%*********************************************************************
\begin{figure}
  \epsfxsize = 8.3 cm
  \epsfbox{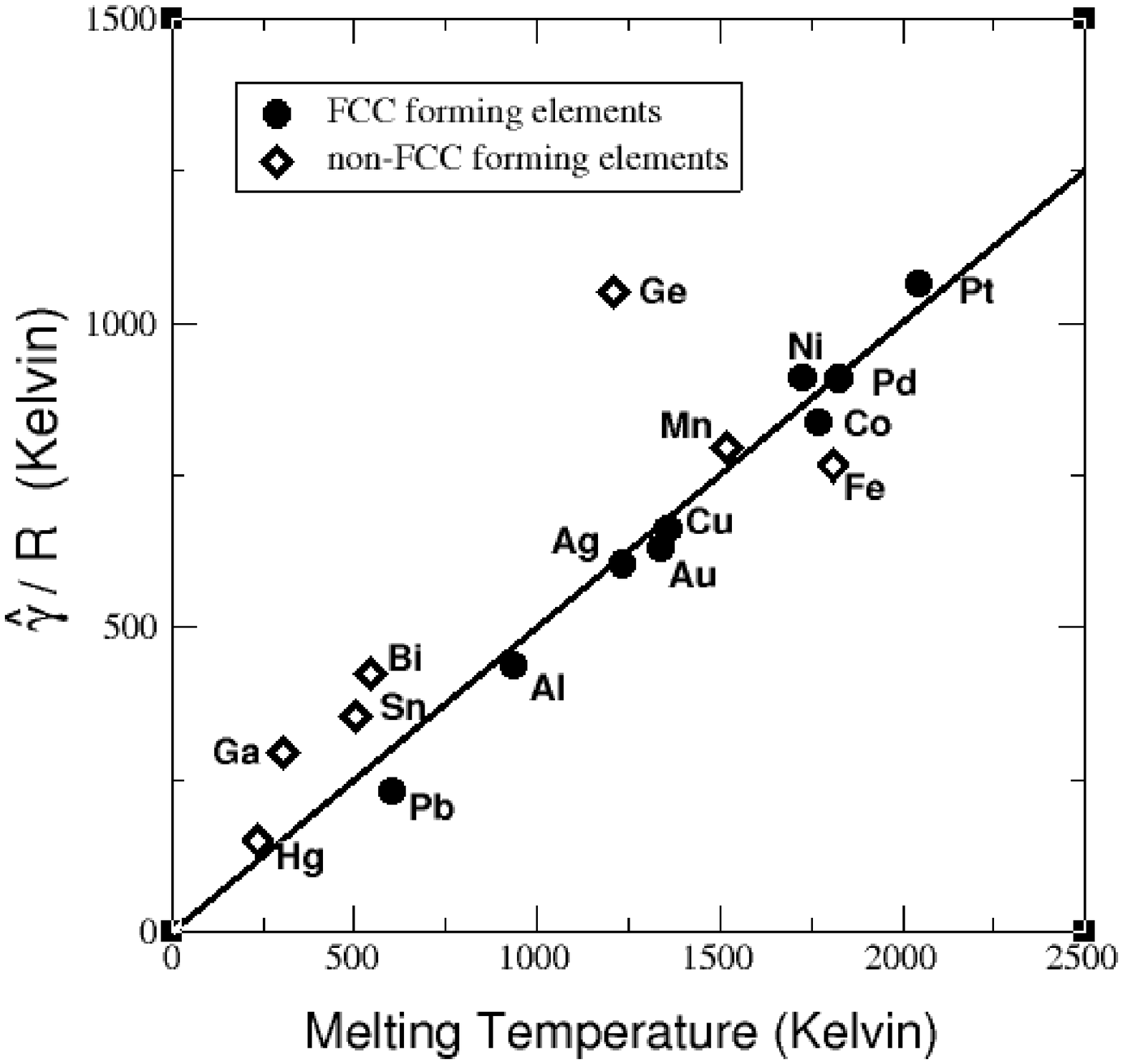}
  \caption{\narrowtext 
Gram-atomic interfacial free energy (scaled by the gas constant to give units of Kelvin) for 
a variety of elemental systems.  Fcc-crystal-forming systems are shown as filled circles and 
values for non-close-packed crystal formers  are shown as open diamonds. The dashed line is a 
line of slope 0.50 representing the best linear fit passing through the origin to the data 
for the fcc crystal formers.}
\label{fig:gamma_vs_t}
\end{figure}
\noindent 
\fi
To test this hypothesis, we plot in Fig.~\ref{fig:gamma_vs_t}, $\hat{\gamma}/R$ as a function 
of melting temperature for a variety of elemental materials using Turnbull's original data.  
Although all of the data exhibits a correlation with $T_m$, the correlation for the 
fcc-crystal forming metals is linear with a slope of 0.50, whereas that for non-fcc forming 
materials exhibits significant scatter. This value of the slope is about 20\% below that 
predicted by 
the direct hard-sphere results, but less than 10\% below that based on the nucleation rate 
result, which is probably a more relevant comparison, since Turnbull's data was also so 
obtained. That the attractive forces contribute only about 10\% to the interfacial free 
energy is consistent with results from simulation\cite{Broughton86,Davidchack00} and 
density-functional theory\cite{Curtin89}
for the Lennard-Jones system, a prototypical fcc-forming model potential. 
It should be noted that Turnbull also reported\cite{Turnbull50} a correlation of $\hat{\gamma}$ with $T_m$, 
but, due to the scatter in the overall data, it was rejected as the basis for an empirical 
rule in favor of the correlation with the enthalpy of fusion.  However, he did observe that 
the correlation with $T_m$ was sensitive to the "complexity" of the crystal structure. This 
is evident in the data shown in Fig.~\ref{fig:gamma_vs_t}, which shows a strong linear correlation with $T_m$, 
for fcc-forming metals, whereas that for the non-close-packed materials is much weaker. The 
hard-sphere interaction does not have the long-range forces necessary to mechanically 
stabilize a non-close packed crystal structure. As a consequence, the interfacial 
thermodynamics of systems that freeze into open crystal lattices will not be well described 
by a purely entropic model and a simple linear scaling with the melting temperature 
is not expected. 

For fcc- (and probably hcp-) forming materials, it is relatively straightforward now to 
understand the empirical correlation of the interfacial free energy with the heat of fusion. 
For the close-packed systems studied by Turnbull, the entropy of fusion, $\Delta_{fus}S$, 
is very nearly constant with an average value very close to that for the hard-sphere system, 
where $\Delta_{fus}S$  = 9.7 J/(mole K). At the melting point, equilibrium requires that  
$\Delta_{fus}H  = T_m \Delta_{fus}S$, so the enthalpy of fusion should scale nearly linearly 
with the melting temperature for these systems. So if $\Delta_{fus}H$ scales approximately 
linear with $T_m$ and $\hat{\gamma}_{hs}$ is proportional to $T_m$ , it then follows that 
$\hat{\gamma}_{hs}$  will exhibit strong linear correlation with $\Delta_{fus}H$, and thus 
Turnbull's rule obtains.

\noindent
{\bf Acknowledgements:}  The author gratefully acknowledges Dr. Ruslan Davidchack for helpful 
conversations and the National Science Foundation for generous under grant CHE-9970903.
%*********************************************************************

\ifpreprintsty
\newpage
\centerline{\bf \Large Figure Captions}
\vskip 0.6cm
\begin{itemize}
\item
{\bf Figure 1:} Gram-atomic interfacial free energy (scaled by the gas constant to give units 
of Kelvin) for a variety of elemental systems.  Fcc-crystal-forming systems are shown as 
filled circles and values for non-close-packed crystal formers  are shown as open diamonds. 
The dashed line is a line of slope 0.50 representing the best linear fit passing through 
the origin to the data for the fcc crystal formers.
\end{itemize}
\newpage
%**********************************************************************
%                           FIGURE 1 CAPTION
%**********************************************************************
\begin{figure}
  \epsfbox{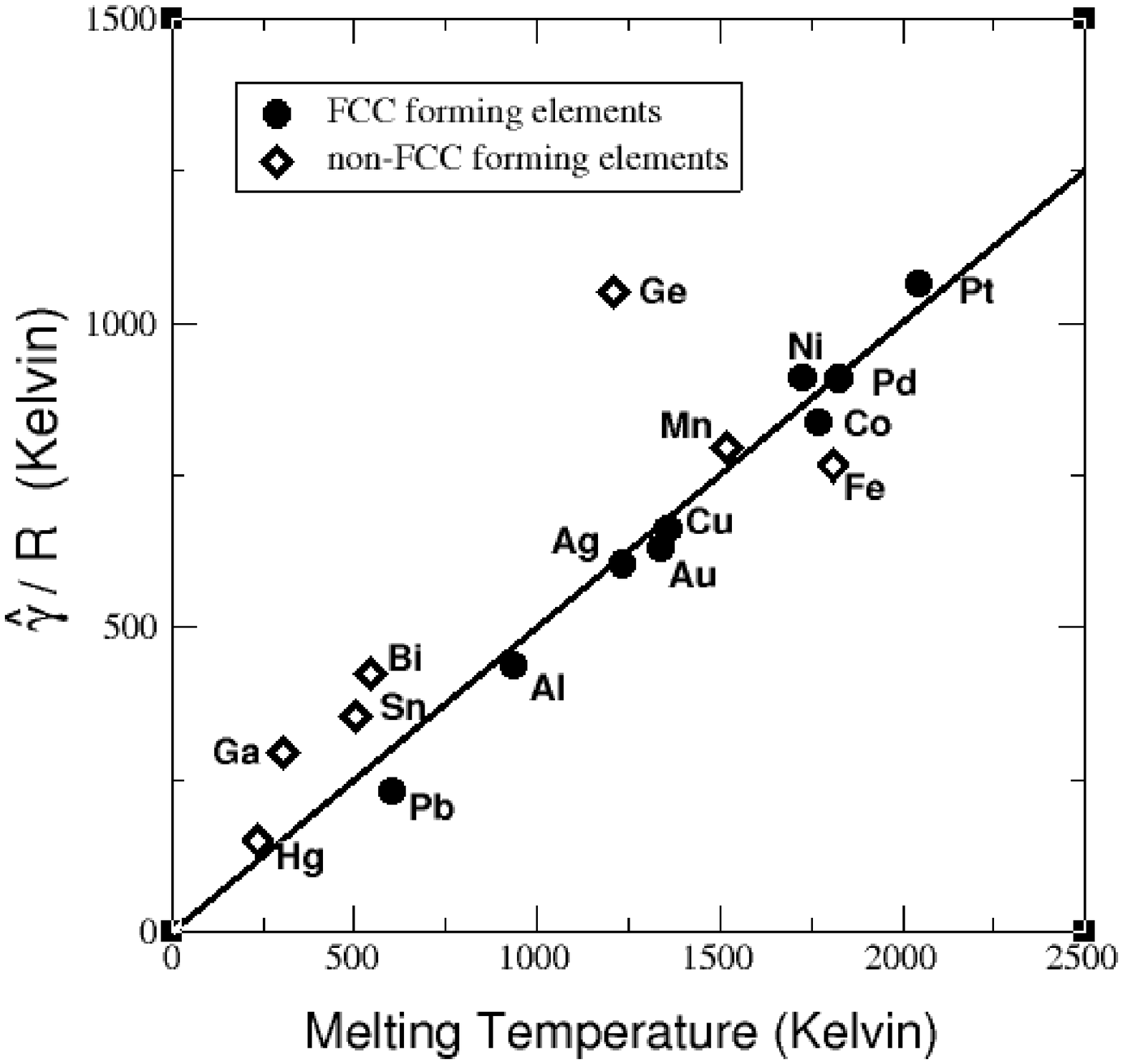}
  \caption{}
\label{fig:gamma_vs_t}
\end{figure}

%**********************************************************************

\else
\end{multicols}
\fi

\end{document}